\documentclass[12pt]{article}

\usepackage[margin=1in]{geometry}
\usepackage{graphicx}
\usepackage{amsmath}
\usepackage{amssymb}
\usepackage{hyperref}
\usepackage{authblk}

\usepackage{subfig}

\usepackage{tikz}
\usepackage{epstopdf}

\title{Coefficient of restitution of a linear dashpot on a rigid surface}

\author[]{Sean P.~Bartz\thanks{sean.bartz@indstate.edu}}
\affil{\emph{Dept.~of Chemistry and Physics, Indiana State University}\\ 
\emph{Terre Haute, IN 47809}}
\graphicspath{{Figures/}}

\begin{document}

\maketitle

\begin{abstract}

The linear dashpot model is applied to a single ball bouncing on a rigid surface. It is shown that when gravity is included the coefficient of restitution depends on impact velocity, in contrast to previous work that ignored the effects of gravity. This velocity dependence is most pronounced at low impact velocities and high damping. 
Previous work has considered the ball to be in contact with the floor when the compression is nonzero, while other analysis terminates the collision earlier, to prevent an attractive force. We compare these models and propose a hybrid between the two.
The hybrid model is successful in reproducing experimental results for a cart bouncing repeatedly on a spring.
\end{abstract}

\section{Introduction}
A spring with a linear dashpot is a convenient model for inelastic collisions, such as those involving viscoelastic spheres.
For the duration of the contact between the spheres, the equations of motion are the well-known damped harmonic oscillator.
In the absence of gravity, this model yields a coefficient of restitution that is independent of impact velocity \cite{Nagurka, schwager_coefficient_2007},  which simplifies the analysis of collisions.
In contrast to a damped harmonic oscillator, a viscoelastic sphere always exhibits a repulsive force. Ensuring the absence of an attractive force in the linear dashpot model affects the calculated collision duration and coefficient of restitution \cite{schwager_coefficient_2007,muller_patric_two-ball_2011}. 

Gravity can be neglected when two spheres collide in midair, as gravity affects each sphere equally. 
However, for a ball bouncing off of a rigid surface, gravity cannot be neglected, and its inclusion results in a coefficient of restitution that depends on impact velocity \cite{Villegas2020}.

There are several experimental methods for determining the elastic and damping properties of a ball, including measuring successive bounce heights \cite{Amrani_2010}, measuring velocities before and after collisions, or measuring the forces directly \cite{cross1999}. Of particular interest to our analysis is the method of measuring the time of flight between successive impacts \cite{Bernstein1977,Nagurka2002, Leconte2006}, the duration of an individual impact \cite{Stensgaard2001} or the total time and number of bounces until the ball comes to rest \cite{Nagurka}.
These experimental techniques typically assume a constant coefficient of restitution, but we show that the final bounces are notably affected by the velocity dependence of $\varepsilon$.
We compare our results to previous theoretical models using the linear dashpot, and experimental data from a cart bouncing on a spring.

\section{Linear dashpot force models}
The spring with a linear dashpot includes a repulsive elastic term that is proportional to the compression and a dissipative term that is proportional to the velocity. 
The compression is defined
$\xi=\min[0,R-z],$
where $R$ is the radius of the ball and $z$ is its vertical position. 
The linear dashpot force is defined
\begin{equation}
F=-k \xi -\gamma \dot{\xi}.\label{eq:forcedef}
\end{equation}
%The equation of motion for the ball is 
%\begin{equation}
%m\ddot{x}=-mg+F.
%\end{equation}
When this force is non-zero, the equation of motion is
\begin{equation}
\ddot{\xi}=-g -\omega_0^2 \xi -2\beta \dot{\xi}, \label{eqEOM}
\end{equation}
with the substitutions 
\begin{equation}
\beta=\frac{\gamma}{2m}, \quad \omega_0^2 =\frac{k}{m}, \quad \omega = \sqrt{\omega_0^2-\beta^2}.
\end{equation}

%The dimensionless solution is
%\begin{eqnarray}
%x(\tau) &=& x_1+x_2 \\
%x_1 &=& \frac{-\dot{x}(0)}{\sqrt{1-\zeta^2}}e^{-\zeta \tau} \sin(\sqrt{1-\zeta^2}\tau) \\
%x_2 &=& -1 +e^{-\zeta \tau}    \left(  \cos\left(\sqrt{1-\zeta^2}\tau\right) +\frac{\zeta}{\sqrt{1-\zeta^2}}\sin \left(\sqrt{1-\zeta^2}\tau \right) \right)
%\end{eqnarray}

Solving for initial conditions $\xi(0)=0, \dot{\xi}(0)=-v_0$ in the case of low damping ($\beta<\omega_0$) gives
%\begin{eqnarray}
%\xi(t) &=& \xi_1+\xi_2 \label{eq:soln1}\\
%\xi_1 &=& -\frac{v_0}{\omega} e^{-\beta t} \sin \omega t \\
%\xi_2 &=& \frac{mg}{k} \left(-1+ e^{-\beta t} \left( \cos \omega t + \frac{\beta}{\omega} \sin \omega t \right) \right),\label{eq:soln3}
%\end{eqnarray}
\begin{equation}
\xi=-\frac{v_0}{\omega} e^{-\beta t} \sin \omega t +
 \frac{mg}{k} \left(-1+ e^{-\beta t} \left( \cos \omega t + \frac{\beta}{\omega} \sin \omega t \right) \right). \label{eq:soln}
\end{equation}
When neglecting gravity, only the first term remains.
%where $\xi_1$ is the homogeneous solution to (\ref{eqEOM}), and $\xi_2$ is the inhomogeneous portion.
In the case of high damping, $\beta>\omega_0$, the solution to (\ref{eqEOM}) is 
%\begin{eqnarray}
%\xi(t) &=& \xi_1+\xi_2 \\
%\xi_1 &=& -\frac{v_0}{\Omega} e^{-\beta t} \sinh \Omega t \\
%\xi_2 &=& \frac{mg}{k} \left(-1+ e^{-\beta t} \left( \cosh \Omega t + \frac{\beta}{\Omega} \sinh \Omega t \right) \right),
%\end{eqnarray}
\begin{equation}
\xi= -\frac{v_0}{\Omega} e^{-\beta t} \sinh \Omega t +\frac{mg}{k} \left(-1+ e^{-\beta t} \left( \cosh \Omega t + \frac{\beta}{\Omega} \sinh \Omega t \right) \right),
\end{equation}
with $\Omega =\sqrt{\beta^2-\omega_0^2}$. 
The following analysis focuses on the low damping case, but the high damping results are found analogously. High damping results are included in the numerical analysis when appropriate.

To examine universal behavior, we define a characteristic velocity $v^*=g\sqrt{k/m}$, and plot the dimensionless quantity $\tilde{v}=v/v^*$. For the sake of comparison, the characteristic velocities for a ping pong ball and a basketball are 0.01 m/s and 0.03 m/s, respectively. 
For the spring-loaded cart experiment \cite{Villegas2020}, $v^*=0.129$ m/s.

\section{Coefficient of restitution}

The coefficient of restitution is defined
\begin{equation}
\varepsilon=\left| \frac{v_f}{v_0} \right|,
\end{equation}
where $v_0$ is the impact velocity and $v_f$ is the velocity at the end of the collision between the ball and the floor. 

Models differ in the definition of the contact duration, and thus have different values of $\varepsilon$. One class of models terminates the collision based on position, when $\xi=0$, which we call models of type I. Alternatively, the end of the collision can be defined when the spring-dashpot force (\ref{eq:forcedef}) becomes zero, which we term model II. 
Additionally, gravity may be included or neglected, which we label with subscripts $a$ and $b$, respectively. 
In all cases, we solve for the contact duration $\tau$, and calculate the coefficient of restitution 
\begin{equation}
\varepsilon=\left | \frac{\dot{\xi}(\tau)}{v_0} \right|.
\end{equation}
The models are compared below. 
%\begin{table}
%\begin{center}
%\begin{tabular}{ c|c|c } 
% 
%   & No gravity & Includes gravity  \\ 
%   \hline
% Collision terminates $\xi=0$ &  model Ia & model Ib \\ 
% Collision terminates $F=0$ & model IIa & model IIb \\ 
%
%\end{tabular}
%
%\end{center}
%\caption{Hhi}
%\end{table}
%
%
%
%\begin{table}
%\begin{center}
%\begin{tabular}{r|c|c|c|c|c}
%
%Model & A & B & C & D& Hybrid\\
%\hline
%Includes gravity & No & Yes & No & Yes & Yes \\
%Collision ends when & $\xi =0$&  $\xi =0$ & $F=0$ & $F=0$ & Hybrid
%\end{tabular}
%
%\end{center}
%
%\end{table}

\subsection{Contact defined by position \label{sec:position}}
Analyses such as \cite{Nagurka, Antypov2011, Sherif2019} consider the ball to be in contact with the floor as long as the compression is nonzero. 
In the zero-gravity case, which we will designate model Ia, the contact duration is calculated \cite{Nagurka}
\begin{equation}
t_{Ia}=\frac{\pi}{\omega} \label{eq:timeIa}
\end{equation}
and the coefficient of restitution is
\begin{equation}
\varepsilon=e^{-\beta t_{Ia}},
\end{equation}
independent of impact velocity.

%In \cite{Villegas2020} an analytic approximation for the coefficient of restitution is found in the case $\beta/\omega_0 \ll 1$
%\begin{equation}
%\varepsilon = e^{-\beta t_{Ib}}\left( \cos \omega t_{Ib} + \frac{g \omega}{v_0 \omega_0^2} \sin \omega t_{Ib} \right) + \frac{\beta g}{v_0 \omega_0^2}, 
%\end{equation}
%where $t_{Ib}$ is the first non-zero solution the  (\ref{eq:soln}) with $\xi=0$. As this equation is transcendental, an  analytic expression for $\varepsilon$ is not achieved.

Including gravity yields an impact velocity-dependent  coefficient of restitution. 
We  numerically solve (\ref{eq:soln}) with $\xi=0$ for the contact time $t_{Ib}$, with sample results shown in Figure \ref{fig:time}. 
 Figures \ref{fig:epsilon}, \ref{fig:epsilon2}, and \ref{fig:epsilon3} show the resulting $\varepsilon$ for a range of $\beta/\omega_0$ for representative impact velocities.

%Assuming that the ball follows the trajectory (\ref{eq:soln}) for the duration of contact gives the unphysical result that the spring-dashpot force becomes attractive near the end of the collision.

\subsection{Contact defined by force}
It has recently been shown that the models in Section \ref{sec:position} result in an unphysical attractive spring-dashpot force near the end of the collision \cite{schwager_coefficient_2007}. This is avoided by defining the end of the collision when the linear dashpot force is equal to zero, which occurs before the compression returns to zero.

We examine this model in the absence of gravity. This calculation is relevant for two colliding spheres that are both moving under the influence of gravity,  and has also been used as an approximation for bouncing on a rigid surface \cite{schwager_coefficient_2007, muller_patric_two-ball_2011}. 
\begin{equation}
t_{IIa}=\frac{1}{\omega}\arccos\left(\frac{2\beta^2}{\omega_0^2}-1\right), 
\label{oldtime}
\end{equation}
and the coefficient of restitution is
\begin{equation}
\varepsilon =e^{-\beta t_{IIa}},
\end{equation}
independent of impact velocity.

\begin{figure}
\includegraphics[width=\textwidth]{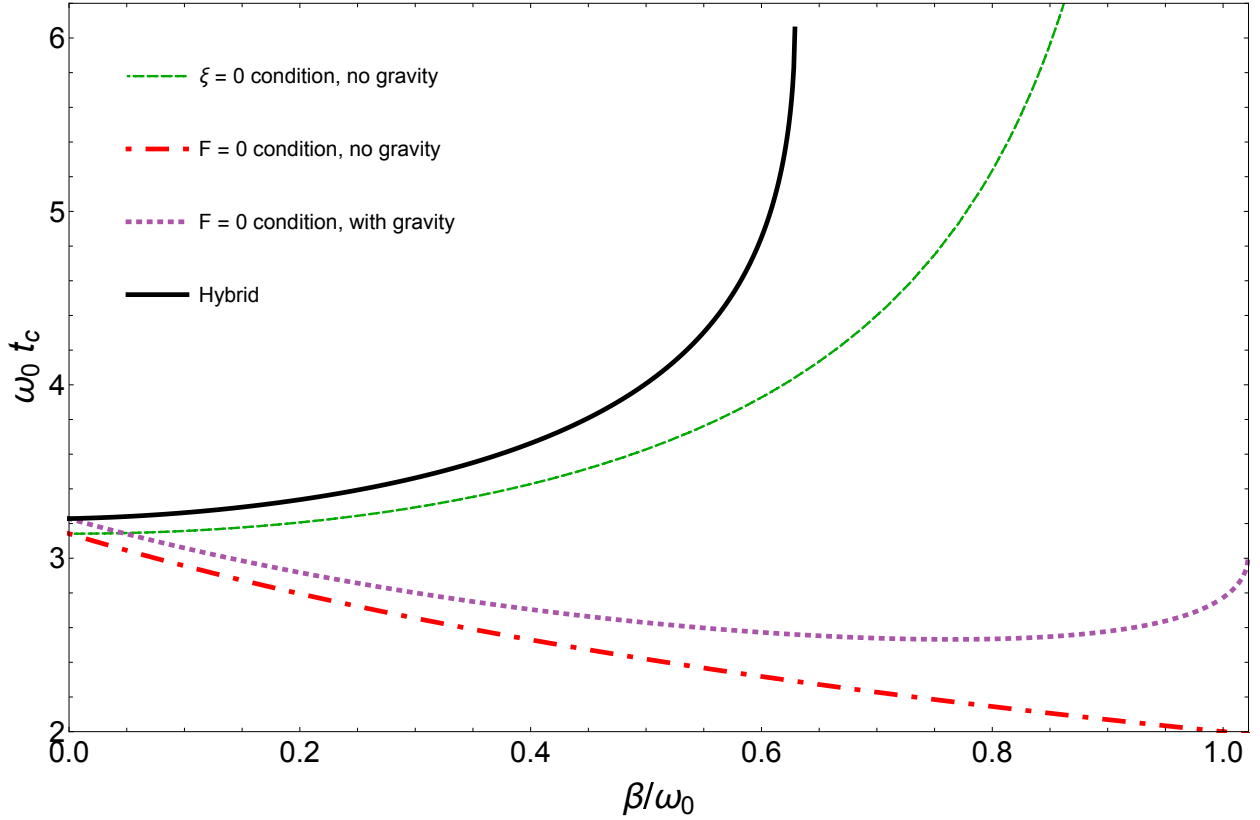}
\caption{The contact time (dimensionless units) is plotted against damping ratios for $\tilde{v}=23$. For models that terminate the collision when compression $\xi=0$ and for the hybrid model, the contact time represents the duration of nonzero compression. For the force based models, the contact time represents the duration of nonzero force. 
The results of model Ib are not shown, as they are visually indistinguishable from the hybrid model. \label{fig:time} }
\end{figure}

Including gravity in the analysis, we solve (\ref{eq:soln}) for the case $\ddot{\xi}(t_{IIb})=-g$, and the ball's motion reverts to free fall. This produces a transcendental equation for $t_{IIb}$
\begin{equation}
-g=v_0 e^{-\beta t_{IIb}} \left(2\beta \cos \omega t_{IIb} +\frac{\omega^2-\beta^2}{\omega} \sin \omega t_{IIb} \right) + ge^{-\beta t_{IIb}} \left(\frac{\beta}{\omega}\sin \omega t_{IIb} - \cos \omega t_{IIb} \right).\label{transcendental}
\end{equation}
This equation is solved numerically for $t_{IIb}$, with sample results shown in Figure \ref{fig:time}. This time is used to calculate the velocity when the force goes to zero, and the coefficient of restitution is calculated based on this rebound velocity.

\begin{figure}
\includegraphics[width=\textwidth]{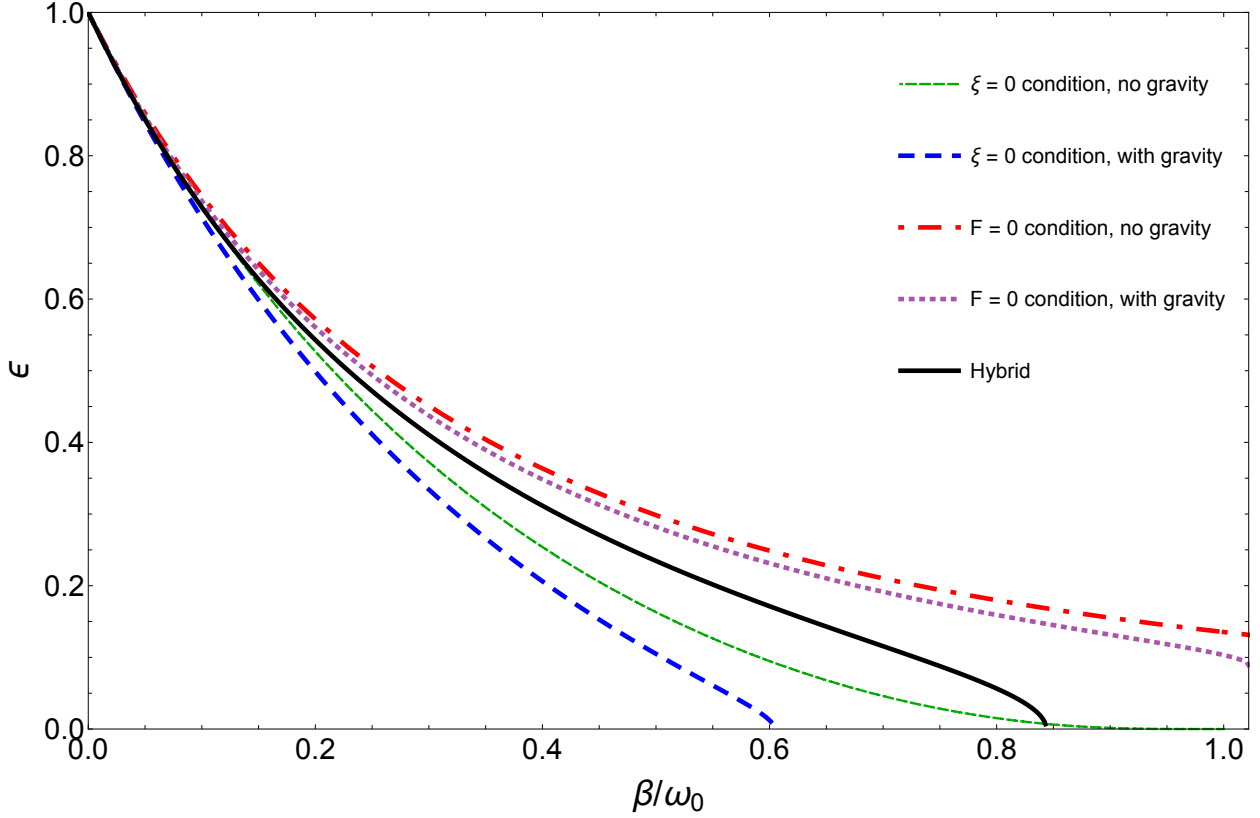}
\caption{The coefficient of restitution for the various models is plotted as a function of ball parameter $\beta/\omega_0$ for a dimensionless impact velocity $\tilde{v}=23$. The curves terminate where elastic capture occurs. \label{fig:epsilon}}
\end{figure}

The coefficient of restitution is found by computing the velocity of the ball at time $t_{IIb}$.
Typical results are shown in Figures \ref{fig:epsilon}, \ref{fig:epsilon2}, and \ref{fig:epsilon3} for a range of damping ratios $\beta/\omega_0$. The coefficients of restitution are strictly less than the result from \cite{schwager_coefficient_2007}.

To compare the contact times between these two models, we calculate the acceleration in the model including gravity, using the contact time from the gravity-free model
\begin{equation}
\ddot{\xi}(t_{IIa})= g e^{-\beta t_{IIa}}.
\end{equation}
This acceleration is greater than zero, indicating that the ball is still in contact with the floor. This shows that the contact duration is greater when gravity is included, as seen in Figure \ref{fig:time}.

\subsection{Hybrid model}
Instantaneous velocities are difficult to measure, so experiments typically infer coefficients of restitution from bounce heights or time between impacts. \cite{Falcon1998}. Doing so implies that the impact velocity and rebound velocity are both measured when the compression is zero, and the ball's position is $z=R$. This assumption is compatible with models of type I, but causes a discrepancy with models of type II. In these models, $z<R$ when the force goes to zero, so the bounce height is lower than predicted, while the flight time is longer.

%We propose a hybrid model, where a velocity is calculated 
%
%Modeling the behavior of the ball during repeated bounces requires  calculating the impact velocity of the ball on the next bounce, $v_1$. In models of type I, this is identical to the rebound velocity  $v_f$, as the ball is at the same position $z=R$ at the beginning and end of each collision, where $\xi=0$. However, for models of type II, the collision ends when the ball is at a position $z<R$. Thus, the ball  spends more time in its upward free-fall motion than the downward portion, and the impact velocity on the next bounce will be less than the rebound velocity.
%
%Experimentally, the coefficient of restitution is often inferred from the bounce height
%
%One benefit of the compression-based models is  clarity on when the rebound velocity should be measured, namely when $\xi=0$. Alternately, the coefficient of restitution can be determined by measuring successive bounce heights
%\begin{equation}
%\varepsilon = \sqrt{\frac{h_{n+1}}{h_n}}, \label{eq:bounceheights}
%\end{equation}
%which assumes that the impact and rebound velocities are both measured at $z=R$. However, in model IIb, the rebound velocity is determined when the compression $\xi<0$, so the ball's position $z<R$, translating to a  lower rebound height. 

\begin{figure}
\includegraphics[width=\textwidth]{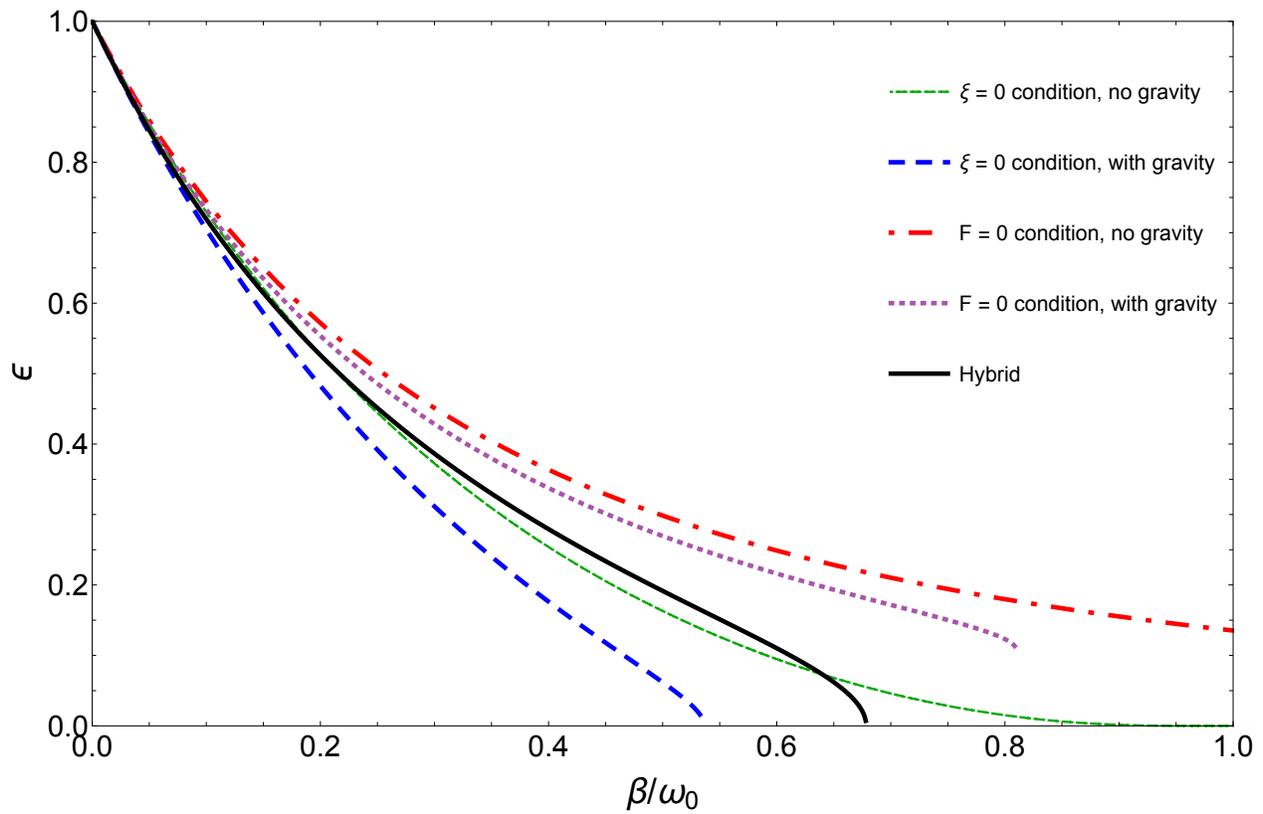}
\caption{The coefficient of restitution is plotted as a function of ball parameter $\beta/\omega_0$ for a dimensionless impact velocity $\tilde{v}=15$. The curves terminate where elastic capture occurs. \label{fig:epsilon2}}\end{figure}

We present a hybrid model, which avoids unphysical attractive forces, but calculates the rebound velocity $v_1$ when the compression is zero.
% which is useful in calculating the rebound height and the impact velocity of the next bounce. 
We calculate the ball's position $z=R-\xi(t_{IIb})$ and velocity $v_f=\dot{\xi}(t_{IIb})$ at the time the force goes to zero, based on (\ref{transcendental}). Afterward, the ball is in free-fall, and we use energy conservation to calculate the velocity when $z=R$
%\begin{equation}
%\frac{1}{2}m v_f^2 = \frac{1}{2}m v_1^2 +mg\xi(t_{IIb})
%\end{equation}
\begin{equation}
v_1=\sqrt{v_f^2-2g\xi(t_{IIb})}, \label{eq:hybridVelocity}
\end{equation}
and  the time it takes the ball to traverse this distance, $\Delta t=2v_1/g$, which is needed when calculating the time between successive bounces, as in Section \ref{sec:multibounce}. 
The coefficients of restitution for this model fall between models Ib and IIb, as seen in Figures \ref{fig:epsilon}-\ref{fig:epsilon3}. 
%The time $t_{IIb}+\Delta t$, plotted in Figure \ref{fig:time}, is visually indistinguishable from that calculated by model Ia, 

\subsection{Inelastic capture}
Inelastic capture occurs when two objects remains in contact after a collision. In model Ia, inelastic capture occurs for $\beta/\omega_0 \geq 1$, as the contact time (\ref{eq:timeIa}) diverges. 
In model IIa, inelastic capture never occurs, as the contact time (\ref{oldtime}) is always finite.

\begin{figure}
\includegraphics[width=\textwidth]{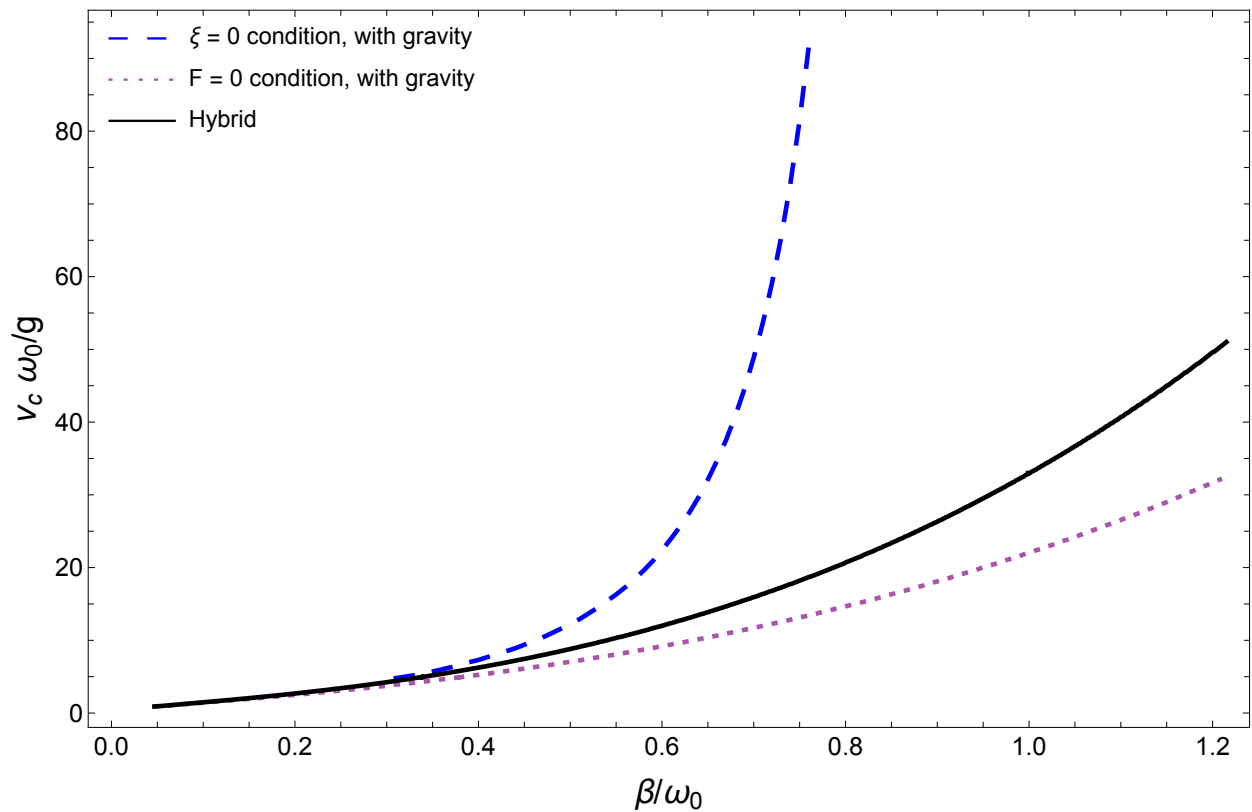}
\caption{For a range of damping ratios $\beta/\omega_0$, the cutoff velocity is plotted, scaled by $g/\omega_0$ to make it dimensionlesss. Below the cutoff  velocity, the ball is  inelasticically captured. \label{fig:critical} }
\end{figure}

In the models that include gravity, there is a critical impact velocity for a given value of $\beta/\omega_0$, below which the ball is inelastically captured. This impact velocity is found when there is no real solution to the collision time (\ref{eq:soln}) and (\ref{transcendental}), respectively. 
In the hybrid model, the critical impact velocity is found when the rebound velocity (\ref{eq:hybridVelocity}) goes to zero.
These critical velocities are shown in Figure \ref{fig:critical}. 
Model IIb and the hybrid model allow for bouncing in the high damping case, but in model Ib, inelastic capture occurs for all velocities at high damping.

\begin{figure}
\includegraphics[width=\textwidth]{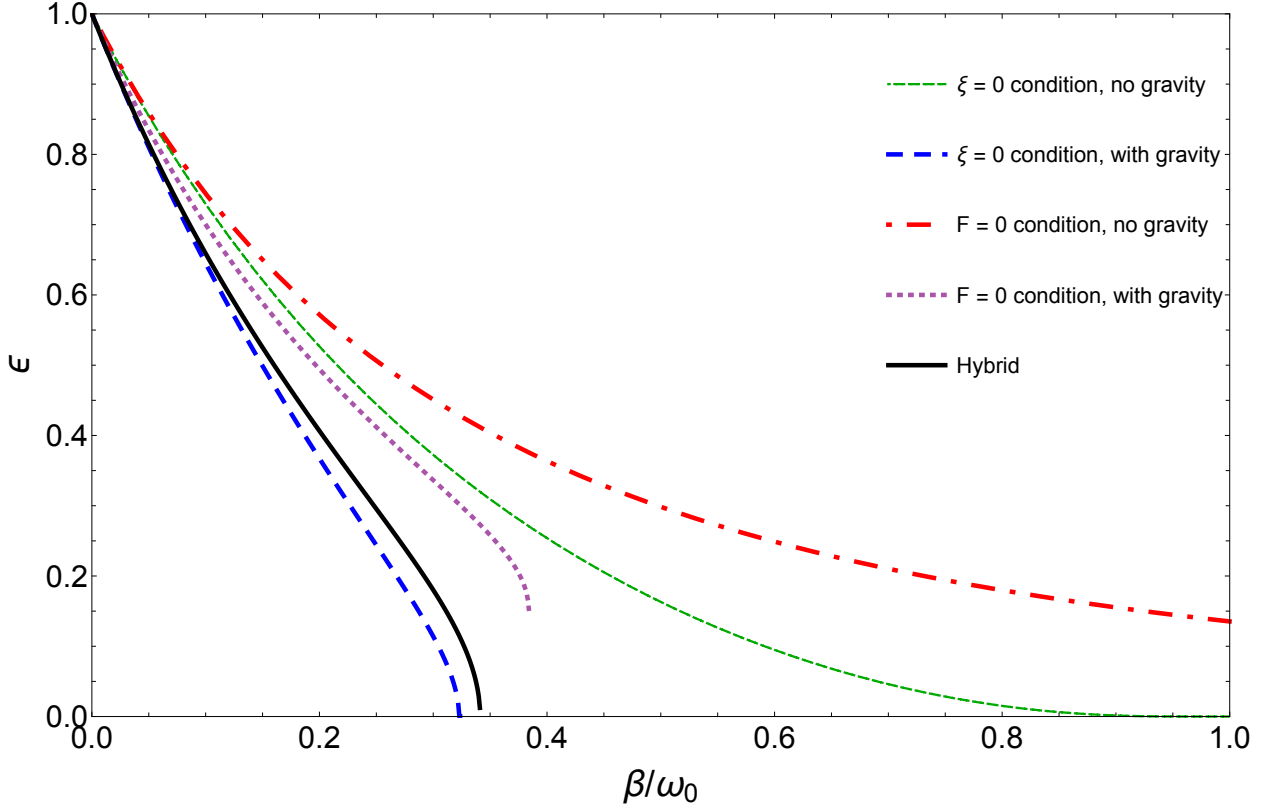}
\caption{The coefficient of restitution is plotted as a function of ball parameter $\beta/\omega_0$ for a dimensionless impact velocity $\tilde{v}=5$. The curves terminate where elastic capture occurs. \label{fig:epsilon3}}
\end{figure}

At the critical impact velocity, $\varepsilon \rightarrow 0$ in model Ib and the hybrid models, as seen in Figures \ref{fig:epsilon}-\ref{fig:epsilon3}, but $\varepsilon$ in model IIb terminates at a finite value. 
This suggests the hybrid model is a more sensible method to remove the unphysical attractive force from the linear dashpot models.

\section{Modeling experimental data \label{sec:multibounce}}
The effects of the velocity-dependent coefficient of restitution are most clearly seen in experiments where an object is allowed to bounce multiple times until  inelastic capture. In \cite{Villegas2020}, the authors used a spring-loaded cart on an inclined track as a practical model of a spring with a linear dashpot. We use the data from that experiment to test the various models

The mass and elastic coefficient are typically measured directly, but the damping coefficient $\beta$ must be estimated from the bouncing motion. In this experiment, $m=0.506$ kg and $k=255$ N/m. The authors use $\varepsilon$at high velocity and the results of model Ia  to estimate $\beta=1.79$ s$^{-1}$. 

\begin{figure}
\includegraphics[width=\textwidth]{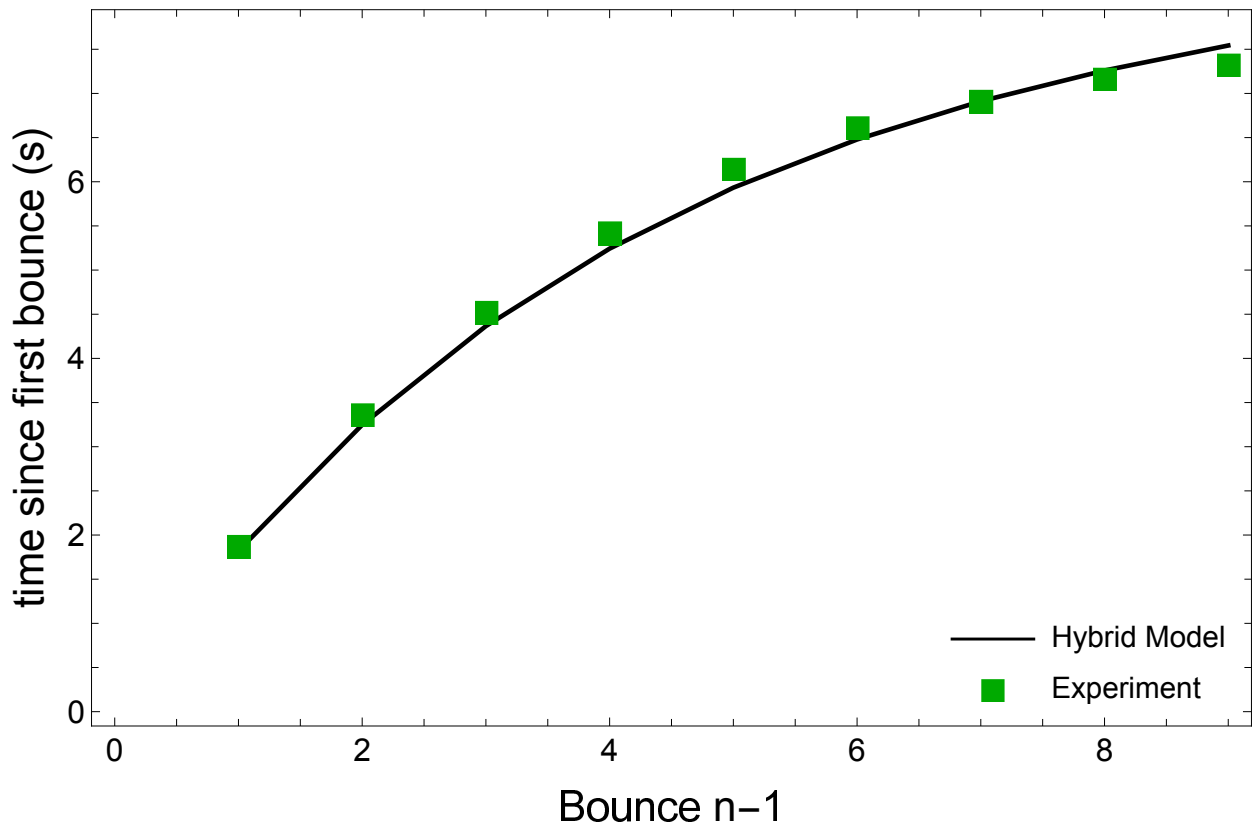}
\caption{The results from the various models are visually indistinguishable on the plot, so only the hybrid model is shown. \label{fig:bounceTime} }
\end{figure}

We use each model recursively,  solving for the successive impact velocities, and totaling the contact time and flight time between each bounce until inelastic capture. The damping coefficient $\beta$ is determined by a best fit to the bounce time data \cite{Villegas2020}, shown in Figure \ref{fig:bounceTime}. The results from the various models are visually indistiguishable, so only the hybrid model is plotted. The values of $\beta$ differ at the 10\% level as shown in Table \ref{tab:betas}.

\begin{table}
\begin{center}

\begin{tabular}{c|c}
Model & Best fit value of $\beta$ (s$^{-1}$)\\
\hline
A & 1.95 \\
B & 1.78 \\
C & 1.94 \\
D & 1.97\\
Hybrid & 1.87

\end{tabular}
\caption{The damping coefficients for the various models are fit by matching the time interval between bounces to the experimental data, as shown in Figure \ref{fig:bounceTime}. \label{tab:betas} }

\end{center}
\end{table}

%We fit $\beta$ to experiment.
%Hybrid model, $\zeta=0.083274$, $\beta=1.87$ s$^{-1}$
%
%$F=0$ with gravity, $\zeta=0.0875$, $\beta=1.967$ s$^{-1}$
%
%$F=0$ no gravity, $\zeta=0.0863$, $\beta=1.94$ s$^{-1}$
%
%$x=0$ no gravity, $\zeta=0.0868$, $\beta=1.95$ s$^{-1}$
%
%$x=0$ with gravity, $\zeta = 0.0795$, $\beta=1.78$ s$^{-1}$

\begin{figure}
\includegraphics[width=\textwidth]{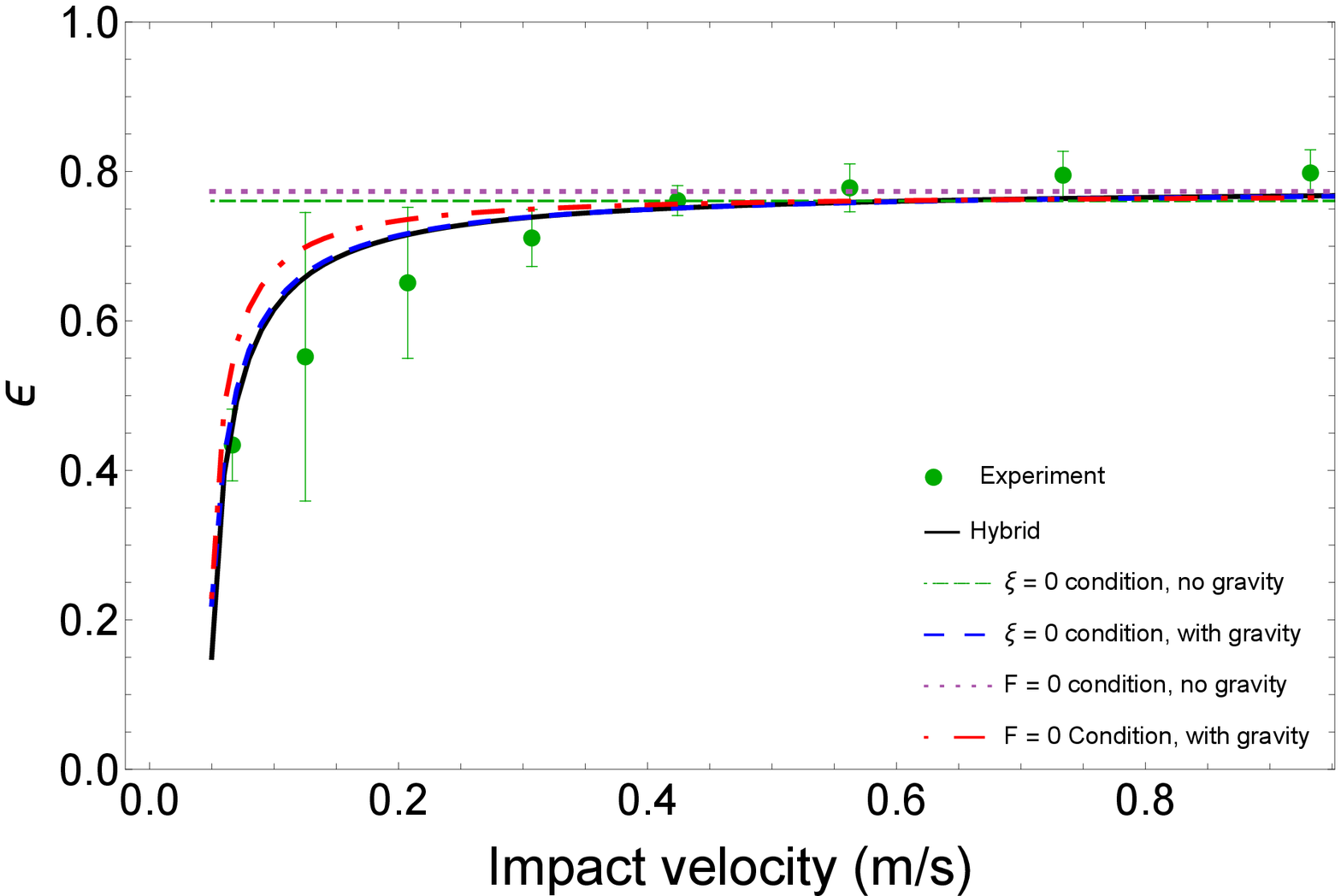}
\caption{The coefficient of restitution $\varepsilon$ is calculated in various models and compared to experimental data from \cite{Villegas2020}. The velocity dependence is seen in the experimental data and well matched by the models that include gravity. \label{fig:epsilonvsVelocity}}
\end{figure}

The differences between models is apparent when these values of $\beta$ are used to calculate $\varepsilon$ for the range of impact velocities in this experiment. Unsurprisingly, the models that account for gravity reflect the observed velocity dependence of the coefficient of restitution, as seen in Figure \ref{fig:epsilonvsVelocity}. The hybrid model and model Ib fit the data equally well. The success of the hybrid model suggests that the rebound velocities were inferred from successive bounce heights, or measured directly when the cart returns to its position of initial impact.


\section{Analysis}
In this paper we compare several spring-dashpot models for a ball bouncing on a rigid surface. We show that the inclusion of gravity produces a velocity-dependent coefficient of restitution, in contrast to the linear dashpot models for colliding spheres. We compare different definitions for the termination of the collision, and their effects on the coefficient of restitution and the contact duration. The differences in these models are most apparent at low damping and low impact velocities.

Comparison to experiment shows these effects. 
For the experiment studied here, the inclusion of gravity has a greater effect on the predictions than the choice of termination condition for the bounce. Although model Ib includes an unphysical attractive force, it gives results just as good as the hybrid model. Absent an independent measurement of $\beta$, these experimental results cannot discriminate between the two models.
The results in Figures \ref{fig:epsilon}, \ref{fig:epsilon2}, and \ref{fig:epsilon3} suggest gravity will also be of increased importance at larger values of $\beta/\omega_0$, and differences between the termination conditions are likely to be observable. 
Collecting data for a cart with a spring with greater damping or for an ``unhappy" ball \cite{Cross2000} and comparing to the predictions of various models would make a good project for an undergraduate student.

We show the importance of relating model choice to the method of experimental measurement. Although model IIb's avoidance of the attractive force is theoretically sound, measuring the rebound velocity at the instant $F=0$ is often impractical. The hybrid model corrects for this difficulty, and is also more accurate in modeling repeated bounces. However, if measuring contact duration with a force sensor, model IIb would be the preferred choice. 

These results suggest that models of collisions of multiple spheres should account for impact velocity dependence when the lower ball is in contact with the floor and gravity is present \cite{muller_patric_two-ball_2011, glendinning_two-ball_2011, berdeni_two-ball_2015, Bartz2022}. These results may also be relevant for modeling granular materials under gravity or another driving force, \cite{Wakou2013} while gas models typically lack such a driving force \cite{McNamara1992} and can neglect this effect.

Observations of a cart bouncing on a spring provide a direct comparison to the linear spring and dashpot model, but bouncing spheres are more accurately described by a  Hertzian force model $F\sim \xi^{3/2}$, with a dissipative force that may also be nonlinear \cite{Antypov2011, Falcon1998}.  These equations of motion cannot be solved analytically, but a student interested in numerical techniques could take a similar approach to the various termination conditions to the one outlined here, and test which  produces more accurate predictions, particularly in cases where the ball's damping coefficient is high.

\bibliography{likeitshot}

\providecommand{\href}[2]{#2}\begingroup\raggedright\begin{thebibliography}{10}

\bibitem{Nagurka}
M.~Nagurka and S.~Huang, \href{http://dx.doi.org/10.23919/ACC.2004.1383652}{``A
  mass-spring-damper model of a bouncing ball,''} in {\em Proceedings of the
  2004 American Control Conference}, vol.~1, pp.~499--504 vol.1.
\newblock 2004.

\bibitem{schwager_coefficient_2007}
T.~Schwager and T.~Pöschel, ``Coefficient of restitution and linear–dashpot
  model revisited,'' \href{http://dx.doi.org/10.1007/s10035-007-0065-z}{{\em
  Granular Matter} {\bfseries 9} no.~6, (Nov., 2007) 465--469},
  \href{http://arxiv.org/abs/0701278}{{\ttfamily arXiv:0701278
  [cond-mat.soft]}}.

\bibitem{muller_patric_two-ball_2011}
{Müller, Patric} and T.~Pöschel, ``Two-ball problem revisited: {Limitations}
  of event-driven modeling,''
  \href{http://dx.doi.org/10.1103/PhysRevE.83.041304}{{\em Physical Review E}
  {\bfseries 83} no.~4, (Apr., 2011) 041304},
  \href{http://arxiv.org/abs/1009.6153}{{\ttfamily arXiv:1009.6153
  [physics.class-ph]}}.

\bibitem{Villegas2020}
C.~E.~P. Villegas, W.~Y. Rojas, C.~Bravo, and A.~R. Rocha, ``Impact dynamics
  for gravity-driven motion of a particle,''
  \href{http://dx.doi.org/10.1088/1361-6404/abb56c}{{\em European Journal of
  Physics} {\bfseries 42} no.~1, (Nov., 2020) 015006}.
  \url{https://doi.org/10.1088/1361-6404/abb56c}.

\bibitem{Amrani_2010}
D.~Amrani, ``Investigating the relationship between the half-life decay of the
  height and the coefficient of restitution of bouncing balls using a
  microcomputer-based laboratory,''
  \href{http://dx.doi.org/10.1088/0143-0807/31/4/002}{{\em European Journal of
  Physics} {\bfseries 31} no.~4, (May, 2010) 717--725}.
  \url{https://doi.org/10.1088/0143-0807/31/4/002}.

\bibitem{cross1999}
R.~Cross, ``The bounce of a ball,''
  \href{http://dx.doi.org/10.1119/1.19229}{{\em American Journal of Physics}
  {\bfseries 67} no.~3, (1999) 222--227},
  \href{http://arxiv.org/abs/https://doi.org/10.1119/1.19229}{{\ttfamily
  https://doi.org/10.1119/1.19229}}. \url{https://doi.org/10.1119/1.19229}.

\bibitem{Bernstein1977}
A.~D. Bernstein, ``Listening to the coefficient of restitution,''
  \href{http://dx.doi.org/10.1119/1.10904}{{\em American Journal of Physics}
  {\bfseries 45} no.~1, (Jan., 1977) 41--44}.
  \url{https://doi.org/10.1119/1.10904}.

\bibitem{Nagurka2002}
M.~L. Nagurka, ``A simple dynamics experiment based on acoustic emission,''
  \href{http://dx.doi.org/https://doi.org/10.1016/S0957-4158(01)00063-0}{{\em
  Mechatronics} {\bfseries 12} no.~2, (2002) 229--239}.
  \url{https://www.sciencedirect.com/science/article/pii/S0957415801000630}.
  Mechatronics Education in Europe and the United States.

\bibitem{Leconte2006}
M.~Leconte, Y.~Garrabos, F.~Palencia, C.~Lecoutre, P.~Evesque, and D.~Beysens,
  ``Inelastic ball-plane impact: An accurate way to measure the normal
  restitution coefficient,'' \href{http://dx.doi.org/10.1063/1.2400061}{{\em
  Applied Physics Letters} {\bfseries 89} no.~24, (Dec., 2006) 243518}.
  \url{https://doi.org/10.1063/1.2400061}.

\bibitem{Stensgaard2001}
I.~Stensgaard and E.~L{\ae}gsgaard, ``Listening to the coefficient of
  restitution -- revisited,'' \href{http://dx.doi.org/10.1119/1.1326077}{{\em
  American Journal of Physics} {\bfseries 69} no.~3, (Mar., 2001) 301--305}.
  \url{https://doi.org/10.1119/1.1326077}.

\bibitem{Antypov2011}
D.~Antypov and J.~A. Elliott, ``On an analytical solution for the damped
  hertzian spring,'' \href{http://dx.doi.org/10.1209/0295-5075/94/50004}{{\em
  {EPL} (Europhysics Letters)} {\bfseries 94} no.~5, (May, 2011) 50004}.
  \url{https://doi.org/10.1209/0295-5075/94/50004}.

\bibitem{Sherif2019}
H.~A. Sherif and F.~A. Almufadi, ``Models for materials damping, loss factor,
  and coefficient of restitution,''
  \href{http://dx.doi.org/10.1115/1.4044281}{{\em Journal of Engineering
  Materials and Technology} {\bfseries 142} no.~1, (Aug., 2019) }.
  \url{https://doi.org/10.1115/1.4044281}.

\bibitem{Cross2000}
R.~Cross, ``The coefficient of restitution for collisions of happy balls,
  unhappy balls, and tennis balls,''
  \href{http://dx.doi.org/10.1119/1.1285945}{{\em American Journal of Physics}
  {\bfseries 68} no.~11, (Nov., 2000) 1025--1031}.
  \url{https://doi.org/10.1119/1.1285945}.

\bibitem{glendinning_two-ball_2011}
P.~Glendinning, ``Two-ball {Newton}'s cradle,''
  \href{http://dx.doi.org/10.1103/PhysRevE.84.067201}{{\em Physical Review E}
  {\bfseries 84} no.~6, (Dec., 2011) 067201}.

\bibitem{berdeni_two-ball_2015}
Y.~Berdeni, A.~Champneys, and R.~Szalai, ``The two-ball bounce problem,''
  \href{http://dx.doi.org/10.1098/rspa.2015.0286}{{\em Proc. R. Soc. Lond. A}
  {\bfseries 471} no.~2179, (July, 2015) 20150286}.

\bibitem{Bartz2022}
S.~P. Bartz, ``Delayed rebounds in the two-ball bounce problem,''
  \href{http://dx.doi.org/10.1088/1361-6404/ac5384}{{\em European Journal of
  Physics} {\bfseries 43} no.~3, (Mar., 2022) 035002}.
  \url{https://doi.org/10.1088/1361-6404/ac5384}.

\bibitem{Wakou2013}
J.~Wakou, H.~Kitagishi, T.~Sakaue, and H.~Nakanishi, ``Inelastic collapse in
  one-dimensional driven systems under gravity,''
  \href{http://dx.doi.org/10.1103/physreve.87.042201}{{\em Physical Review E}
  {\bfseries 87} no.~4, (Apr., 2013) }.
  \url{https://doi.org/10.1103/physreve.87.042201}.

\bibitem{McNamara1992}
S.~McNamara and W.~R. Young, ``Inelastic collapse and clumping in a
  one-dimensional granular medium,''
  \href{http://dx.doi.org/10.1063/1.858323}{{\em Physics of Fluids A: Fluid
  Dynamics} {\bfseries 4} no.~3, (Mar., 1992) 496--504}.
  \url{https://doi.org/10.1063/1.858323}.

\bibitem{Falcon1998}
E.~Falcon, C.~Laroche, S.~Fauve, and C.~Coste, ``Behavior of one inelastic ball
  bouncing repeatedly off the ground,''
  \href{http://dx.doi.org/10.1007/s100510050283}{{\em The European Physical
  Journal B} {\bfseries 3} no.~1, (June, 1998) 45--57}.
  \url{https://doi.org/10.1007/s100510050283}.

\end{thebibliography}\endgroup
\bibliographystyle{utphys}

%%%%%%%
% A Small Purple Boat production.
%     ¯\_(ツ)_/¯     %
%%%%%%%
\end{document}